\documentclass[aps,twocolumn,showpacs,prl,preprintnumbers,amsmath,amssymb]{revtex4}

\usepackage{graphicx}
\usepackage{dcolumn}
\usepackage{bm}

\usepackage[usenames]{color}
\usepackage{ulem}

\begin{document}

\preprint{APS/123-QED}

\title{Vortex Crystals with Chiral Stripes in Itinerant Magnets}

\author{Ryo Ozawa$^1$, Satoru Hayami$^2$, Kipton Barros$^2$, Gia-Wei Chern$^{2, 3}$, Yukitoshi Motome$^1$, and Cristian D. Batista$^2$}
\affiliation{%
$^1$Department of Applied Physics, University of Tokyo, Tokyo 113-8656, Japan\\
$^2$Theoretical Division and CNLS, Los Alamos National Laboratory, Los Alamos, New Mexico 87545, USA\\
$^3$Department of Physics, University of Virginia, Charlottesville, VA 22904, USA 
}%




\date{\today}

\begin{abstract}
We study noncoplanar magnetic ordering in frustrated itinerant magnets. 
For a family of Kondo square lattice models with classical local moments, we find that a double-$Q$ noncoplanar vortex crystal  has  lower energy than the single-$Q$ helical order expected from the Ruderman-Kittel-Kasuya-Yosida interaction whenever the lattice symmetry dictates four global maxima in the bare magnetic susceptibility. 
By expanding in the small Kondo exchange and the degree of noncoplanarity, 
we demonstrate that this noncoplanar state arises from a Fermi surface instability, 
and it is generic for a wide range of electron filling fractions whenever the two ordering wave vectors connect independent sections of the Fermi surface. 
\end{abstract}


\maketitle

Noncoplanar spin textures in itinerant magnets are generating increasing interest 
because of the associated spin Berry phase, 
which induces a tremendous effective magnetic field on the itinerant electrons~\cite{Berry, 
PhysRev.115.485, RevModPhys.82.1959, PhysRevB.45.13544, PhysRevLett.69.3232, PhysRevLett.83.3737}.
The Berry phase is proportional to the local  spin scalar chirality $\chi_{ijk} =  \bm{S}_i\cdot \left(\bm{S}_j  \times \bm{S}_k \right)$, 
i.e., the triple product of neighboring local magnetic moments $\bm{S}_i$,  
and may change the topology of the electronic band structure 
leading to the so-called topological quantum Hall effect~\cite{PhysRevB.62.R6065, taguchi2001spin, binz2008chirality}.

Although noncoplanar spin textures are not common in Mott insulators (local moment systems) with small spin anisotropy, 
they do appear  frequently in itinerant magnets~\cite{PhysRevLett.101.156402, JPSJ.79.083711, PhysRevLett.105.226403, 
PhysRevLett.108.096401, PhysRevLett.109.166405, PhysRevB.90.060402, PhysRevB.90.245119}. 
They are characterized by  multiple ordering wave vectors (multiple-$Q$) 
that maximize the bare magnetic susceptibility by connecting pieces of Fermi surface (FS). 
The noncoplanar orderings reported so far arise only for particular band structures and electronic filling fractions that
maximize the magnetic susceptibility at high-symmetry wave vectors~\cite{smallpoint}.  
Here, we ask if noncoplanar magnetic orderings arise under more general conditions in 
frustrated itinerant magnets.

To address this question, we consider metallic systems whose bare magnetic susceptibility is maximized 
by multiple competing low-symmetry wave vectors ${\bm  Q}_{\nu}$, 
a common situation that arises over a wide range of electronic band fillings. 
The multiplicity of susceptibility maxima indicates frustration, i.e., competition of phases with similar energies.
For concreteness, we study the square Kondo lattice model 
with four susceptibility maxima at  ${\bm  q}=\pm {\bm Q}_{\nu}$ with ${\nu =1,2}$.
Our main result is the discovery of a generic ground state given by a double-$Q$ magnetic ordering. 
Importantly, double-$Q$ order implies a noncoplanar spin configuration.
The state we find is a vortex-antivortex crystal with a one-dimensional modulation (stripes) of spin scalar chirality~\cite{ivar}. 
Large-scale, unbiased Langevin dynamics simulations based upon the kernel polynomial method 
indicate that our double-$Q$ state arises ubiquitously~\cite{PhysRevB.88.235101, numeric}. 
In this Letter, we show that our state has lower energy, for a wide range of electron filling fractions, 
than the single-$Q$ helical coplanar ordering that would naively be inferred from the 
Ruderman-Kittel-Kasuya-Yosida (RKKY) interaction~\cite{PhysRev.96.99,Kasuya01071956,PhysRev.106.893}.
Our unexpected result appears only after careful examination of higher-order terms in a perturbative expansion 
in the Kondo exchange coupling and in the degree of noncoplanarity.  
For certain commensurate wave-vectors, we confirm our perturbative analysis by full diagonalization of the Hamiltonian.
Our results unveil the origin of the ubiquitous noncoplanar orderings in frustrated itinerant magnets 
in the absence of relativistic spin-orbit coupling.

We begin with the Kondo lattice Hamiltonian,
\begin{align}
\!\!\!\!{\cal H} =\!\!\sum_{\bm{k}\in {\rm 1BZ}, \sigma}  \!\!\varepsilon(\bm{k}) c_{\bm{k}\sigma}^{\dagger}c_{\bm{k}\sigma}^{\;} 
- J\!\!\!\!\!\!\!\!\sum_{{\bm k}, {\bm q}\in {\rm 1BZ}, \sigma, \sigma'}  \!\!\!\!\!\!\!\!c_{{\bm k}\sigma}^{\dagger}{\bm \sigma}_{\sigma\sigma'} c_{{\bm k}+{\bm q}\sigma'}^{\;} \cdot {\bm S}_{\bm q},
\label{eq:Ham_KLM_k-sp}
\end{align}
defined on the square lattice. 
The operator $c_{{\bm k}\sigma}^\dagger (c_{{\bm k}\sigma}^{\;})$ creates (annihilates) 
an itinerant electron with momentum ${\bm k}$ and spin $\sigma$. 
${\rm 1BZ}$ denotes the first Brillouin zone (BZ) of the square lattice, and the bare electronic dispersion relation is 
$\varepsilon(\bm k)= \sum_{jl} t_{jl} e^{i {\bm k} \cdot {\bm r}_{jl}}$, where $t_{jl}$ are hopping integrals between sites $j$ and $l$.  
We consider the first- and third-neighbor hoppings,  $t_1$ and $t_3$, respectively,  
although our results hold for generic band structures.  
The second term of ${\mathcal H }$ corresponds to the exchange coupling ($J$) between the conduction electrons 
and the localized magnetic moments ${\bm S}_{\bm q}$ in the Fourier space representation.
We assume classical moments with magnitude $|\bm{S}_i|=1$.   
$\bm \sigma$ is the vector of the Pauli matrices. 
Hereafter, we take $t_1=1$ and $a=1$ (lattice constant) as energy and length units, respectively.
 
We consider the ground state magnetic orderings  induced by the FS instabilities 
in the weak-coupling regime, $J \rho(\varepsilon_{\rm F}) \ll 1$, 
where $\rho(\varepsilon_{\rm F})$ is the density of states at the Fermi level $\varepsilon_{\rm F}$. 
At second order in $J$, the effective spin Hamiltonian is the RKKY model,
\begin{align}
\mathcal{H}^{(2)}_{\rm eff}  = -J^2 \sum_{{\bm q} \in {\rm 1BZ}}\chi^0_{\bm q} \, |\bm{S}_{\bm q}|^2.
\label{eq:RKKY_sus}
\end{align}
Here, $\chi^0_{\bm q} =T\sum_{\bm{k}\in {\rm 1BZ}, \omega_n} G^0_{\bm{k}, \omega_n} G^0_{\bm{k}+\bm{q}, \omega_n}$ 
is the bare susceptibility of the conduction electrons, 
where $T$ is temperature, $\omega_n = (2n+1)\pi T$ are the Matsubara frequencies, 
$G^0_{\bm{k}, \omega_n}= \left\{ \omega_n - \left[\varepsilon ({\bm k}) - \mu\right]\right\}^{-1}$ 
is the bare Green function, and $\mu$ is the chemical potential. 
We note that $\sum_{\bm q} | \bm{S}_{\bm q}|^2 = \sum_{i=1}^N | \bm{S}_{i}|^2 = N$ ($N$ is the number of sites) 
and thus a single-$Q$ helical state whose ordering vector maximizes $\chi^0_{\bm q}$ 
is a ground state of $\mathcal{H}^{(2)}_{\rm eff}$.

\begin{figure}[t!]
	\includegraphics[width=8.0cm]{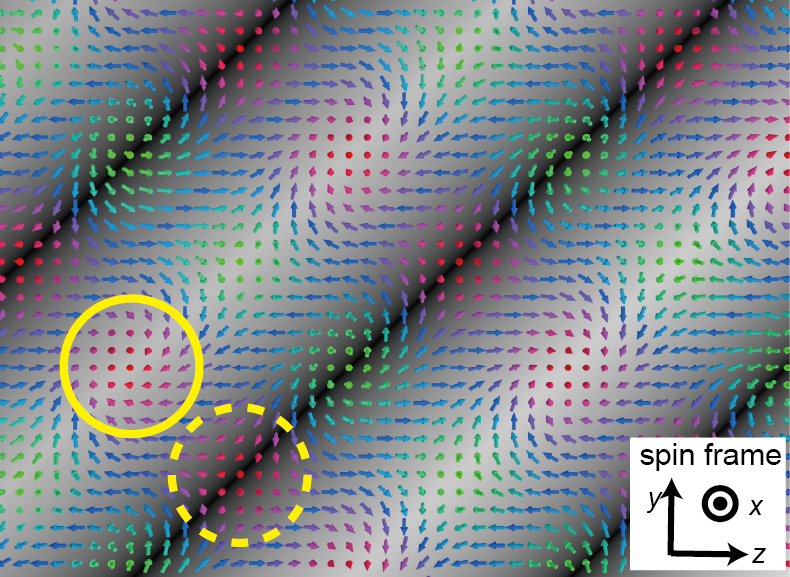}
	\caption{ \label{fig:meron} (color online).
	Vortex crystal with scalar chiral stripes in Eq.~(\ref{eq:Spin_meron}). 
	The arrows represent the in-plane ($yz$) spin component 
	and the colors indicate the out-of-plane ($x$) spin component. 
	The spin frame is rotated from Eq.~(\ref{eq:Spin_meron}) to show the vortex structure clearly.
	The gray-scale background shows striped modulation of the spin scalar chirality. 
	In this example, we chose $\bm{Q}_1=(Q, Q)$, $\bm{Q}_2=(Q, -Q)$ with $Q=\pi/12$, and $b=1$.
	Solid and dashed circles show vortex and antivortex spin patterns. 
	}
\end{figure}
\begin{figure}[t]
	 \includegraphics[width=8.0cm]{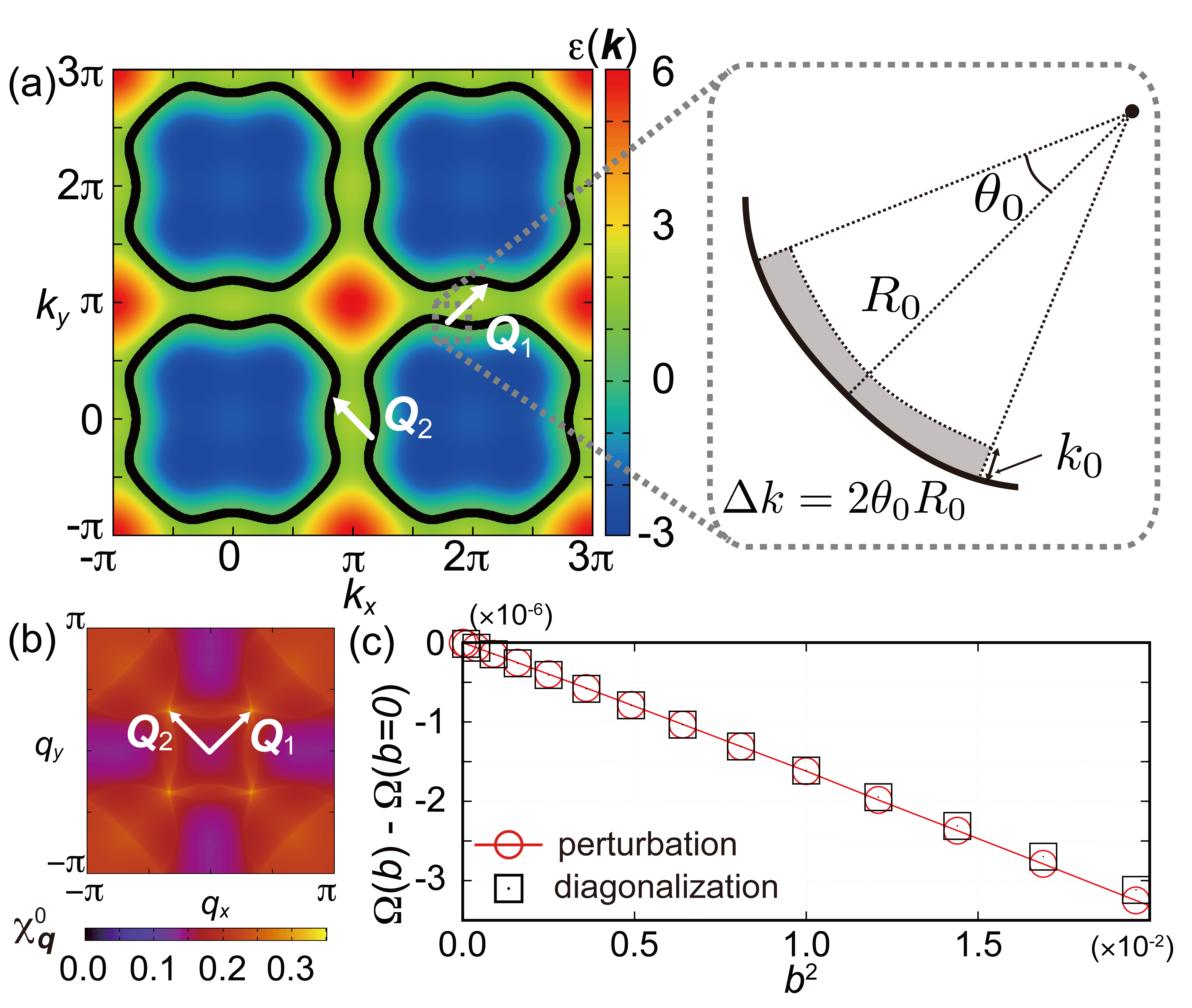}
	 \caption{ \label{fig:perturbation} (color online).
	 (a)  FS (black curves) and energy dispersion $\varepsilon({\bm k})$ (contour plot) 
	 for $t_3=-0.5$, $J=0$, and $\mu=0.98$ in an extended BZ. 
	 ${\bm Q}_1$ and ${\bm Q}_2$ denote the connecting vectors of FS 
	 which maximize $\chi^0_{\bm q}$ shown in (b).
	 In the enlarged figure, we show the way to approximate the FS near the hot spots in the cylindrical coordinate 	 
	 ($R_0$, $\theta_0$) for the perturbation theory (see the text).  
	 (c) Grand potential of vortex crystals with ${\bm Q}_1=(\pi/3, \pi/3)$ as a function of $b^2$  
	 for $t_3=-0.5$, $J=0.1$, and $\mu=0.98$. 
	 The results are calculated for the system size $N=960^2$ sites.  
	 The circles and squares show the results by the perturbative approach 
	 and the exact evaluation by using direct diagonalization, 
	 respectively.
	 }
\end{figure}

When ${\bm  q}={\bm Q}_1$ is a low-symmetric wave vector parallel to the $(\pi,0)$ or $(\pi,\pi)$ direction, 
we have another low-symmetric wave vector, ${\bm Q}_2= R {\bm Q}_1 \neq {\bm Q}_1$, where $R$ is a rotation by $\pi/2$ 
[see Fig.~\ref{fig:perturbation}(b) as an example of ${\bm Q}_1=(\pi/3,\pi/3)$]. 
In this case, the bare magnetic susceptibility has four global maxima at $\pm {\bm Q}_{\nu}$ ($\nu=1, 2$).
In this situation, multiple-$Q$ magnetic orderings obtained from linear superpositions of these four modes 
are anticipated to compete against the single-$Q$ helical ordering. 
We argue that the relevant variational ansatz is a double-$Q$ vortex crystal state,
\begin{align}
\!\!{\bm S}^{\rm vc}_i(b)\!\!=\!\!
	\left(\!\!\begin{array}{c}
	\sqrt{(1-b^2) + b^2\cos^2({\bm Q}_2\cdot{\bm r}_i) } \cos({\bm Q}_1\cdot{\bm r}_i)\\
	\sqrt{(1-b^2) + b^2\cos^2({\bm Q}_2\cdot{\bm r}_i) } \sin({\bm Q}_1\cdot{\bm r}_i)\\
	b\sin({\bm Q}_2\cdot{\bm r}_i)
	\end{array}
	\!\!\right)\!\!,\label{eq:Spin_meron}
\end{align}
where $\bm{r}_i$ is the position vector of the site $i$.
Note that $b = 0$ corresponds to an ordinary helical state in the $xy$ spin-plane with the wave vector $\bm{Q}_1$.
We will show, however, that the system can lower its energy by introducing an out-of-plane ($z$ component) modulation with amplitude $b \neq 0$ and wave vector $\bm{Q}_2$. 
Figure~\ref{fig:meron} shows this vortex crystal state with a maximal degree of noncoplanarity, $b=1$.
Interestingly, this state also exhibits modulation of the scalar chirality~\cite{localchi} along the $\bm{Q}_2$ direction, 
which we illustrate in the gray scale background of Fig.~\ref{fig:meron}.

The spin configuration of the vortex crystal in Eq.~\eqref{eq:Spin_meron} has higher harmonics, 
which arise from an expansion of the square root prefactors introduced to satisfy the normalization condition $|\bm{S}_i|=1$. 
Except for a few particular commensurate wave vectors, this is true in general for any multiple-$Q$ solution. 
This implies that multiple-$Q$ orderings have higher energy than the single-$Q$ helical ordering at the RKKY level. 
However, the balance changes once higher-order terms are included. 
The fourth-order contribution to the perturbative expansion in $J$ gives
\begin{align}
\!\!\!\!\mathcal{H}^{(4)}_{\rm eff} \!=
 -  J^4  \!\!\!\!\!\!\!\!\!\! \sum_{{\bm q}_1+\cdots+{\bm q}_4 ={\bm 0}}
\!\!\!\!\!\!\!\!\! \Pi_{{\bm q}_1, {\bm q}_2, {\bm q}_3, {\bm q}_4} 
\left(\bm{S}_{{\bm q}_1}\!\!\cdot \bm{S}_{{\bm q}_2}\right) 
\left(\bm{S}_{{\bm q}_3}\!\!\cdot \bm{S}_{{\bm q}_4}\right),
\label{eq:general_exp_J4}
\end{align}
where $\Pi_{{\bm q}_1, {\bm q}_2, {\bm q}_3, {\bm q}_4}$ is the coefficient of scattering process 
with ${\bm q}_1, {\bm q}_2, {\bm q}_3$, and ${\bm q}_4$. 
We focus on the dominant contributions denoted by $A_{\bm q}=\Pi_{{\bm q}, -{\bm q}, {\bm q}, -{\bm q}}$,  
$B_{{\bm q}, {\bm q}'}=\Pi_{{\bm q}, {\bm q}', -{\bm q}', -{\bm q}}$, and 
$W_{{\bm q}, {\bm q}'}=\Pi_{{\bm q}, {\bm q}', -{\bm q}, -{\bm q}'}$, whose expressions are given in~\cite{supp}. 
By Fourier transforming the spin configuration in Eq.~\eqref{eq:Spin_meron} and substituting into Eq.~(\ref{eq:general_exp_J4}),
we obtain the energy (grand potential) up to ${\cal O} (J^4)$, assuming that $b \ll 1$:
\begin{align}
&\Omega^{\rm (4)}_{\rm eff}(b) = - J^2\left[\left(1 - \frac{b^4}{32}\right)\chi^0_{{\bm Q}_1} 
+ \frac{b^4}{32}\chi^0_{{\bm Q}_1+2{\bm Q}_2} \right]\nonumber\\
&-\frac{J^4}{2} \!\!\left[(1 - b^2)A_{{\bm Q}_1} + 2b^2 B_{{\bm Q}_1, {\bm Q}_2} - b^2 W_{{\bm Q}_1, {\bm Q}_2}\right].
\label{gp}
\end{align}
The energy difference between the $b\neq 0$ double-$Q$ state and the $b=0$ single-$Q$ state, 
$\Delta \Omega^{(4)}_{\rm eff}(b) \equiv \Omega^{(4)}_{\rm eff}(b) - \Omega^{(4)}_{\rm eff}(b=0)$, is given by
\begin{align}
\Delta \Omega^{(4)}_{\rm eff}(b) =&\frac{J^2b^4}{32}\left[ \chi^0_{{\bm Q}_1} - \chi^0_{{\bm Q}_1 + 2{\bm Q}_2} \right]
\nonumber \\
&+\frac{J^4b^2}{2}\left[ A_{{\bm Q}_1} - 2B_{{\bm Q}_1, {\bm Q}_2} + W_{{\bm Q}_1, {\bm Q}_2}\right]
\nonumber \\
\equiv& \alpha J^2b^4 - \beta J^4b^2.
\label{eq:del_energy_Jb_direct_simple}
\end{align}
The coefficient $\alpha$ is always positive because $\chi^0_{{\bm q}}$ is maximized at ${\bm q}=\pm{\bm Q}_1, \pm{\bm Q}_2$. 
If $\beta$ also turns out to be positive, the optimal value of $b$ is $b_{\rm opt} = \sqrt{\beta/(2 \alpha)} J$.  
Indeed, the explicit evaluation of the coefficients  $A_{{\bm Q}_1}$,  $B_{{\bm Q}_1, {\bm Q}_2}$, and 
$W_{{\bm Q}_2, {\bm Q}_2}$ gives $\beta>0$ at a low enough $T$, as long as ${\bm Q}_{\nu}$ connects the FS. 
Moreover, $A_{{\bm Q}_1}$, $B_{{\bm Q}_1, {\bm Q}_2}$, and $W_{{\bm Q}_1, {\bm Q}_2}$ can diverge for $T\rightarrow 0$.
While this observation implies that the perturbative expansion is not valid in the $T \to 0$ limit, 
it suggests that $b$ can become of order one even for very small $J$. 

To verify this hypothesis, we first derive a regular (nondivergent) perturbative treatment of the instability 
of the single-$Q$ state towards the noncoplanar double-$Q$ ordering described by Eq.~\eqref{eq:Spin_meron}. 
For this purpose, we change the local reference frame for the spin components of the itinerant electrons, 
such that the local quantization $x$-axis is aligned with the local moments of the single-$Q$ helical state 
with ${\bm q} = {\bm Q}_1$ at each site.
The quasi-particle operator in the new reference frame is 
$\tilde{c}^{(\dagger)}_{{\bm k}\sigma} = c^{(\dagger)}_{[{\bm k}+ \sigma{\bm Q}_1/2] \sigma}$.
Then, up to the quadratic order in $b$ (we assume that $b \ll 1$), we obtain the following expression for $\mathcal{H}$
with the local moment configuration given by Eq.~\eqref{eq:Spin_meron}:
\begin{align}
\label{eq:Ham_KLM_b_exp_new_frame}
	\mathcal{H}_{\rm eff}^{\rm vc}(b)
	=& \!\!\!\sum_{{\bm k} \in {\rm 1BZ}, \sigma} \left[
	\tilde{\varepsilon}({\bm k}, \sigma)\tilde{c}^\dagger_{{\bm k}\sigma}\tilde{c}^{\;}_{{\bm k} \sigma} 
	- J\left(1 - \frac{b^2}{4}\right)  \tilde{c}^\dagger_{{\bm k}\sigma}\tilde{c}^{\;}_{{\bm k}{-\sigma}}\right.
	\nonumber\\ 	
	&- i\frac{Jb}{2} \left(
	\tilde{c}^\dagger_{{\bm k}+{\bm Q}_2 \sigma}\tilde{c}^{\;}_{\bm k\sigma} 
     	-\tilde{c}^\dagger_{{\bm k}\sigma}\tilde{c}^{\;}_{{\bm k}+{\bm Q}_2 \sigma}  \right)
 	\nonumber\\
	&\left.+ \frac{Jb^2}{8} \left(\tilde{c}^\dagger_{{\bm k}+2{\bm Q}_2 \sigma}\tilde{c}^{\;}_{\bm k-\sigma} 
     	+\tilde{c}^\dagger_{{\bm k}\sigma}\tilde{c}^{\;}_{{\bm k}+2{\bm Q}_2 -\sigma}  \right)\right],
\end{align}
with $\tilde{\varepsilon}({\bm k}, \sigma) = \varepsilon({\bm k} + \sigma{\bm Q}_1/2)$.
This is a better basis for applying perturbation theory because electronic scattering with the ${\bm Q}_1$ component of
the spin configuration is accounted for exactly via diagonalization of the $2\times2$ matrices associated with 
the first two terms of $\mathcal{H}_{\rm eff}^{\rm vc}(b)$.  
The divergence of the {\it fourth-order} contributions to the perturbative treatment in the original basis is avoided in this new basis 
because they are incorporated at {\it second order} in the amplitude of the last two terms of $\mathcal{H}_{\rm eff}^{\rm vc}(b)$~\cite{supp, 2order}. 

The most significant second-order contributions to the total energy are given by the regions around the two pairs of FS 
points (hot spots) connected by the ordering vectors $\bm{Q}_1$ and  $\bm{Q}_2$ 
[see Figs.~\ref{fig:perturbation}(a) and \ref{fig:perturbation}(b)].
The dispersion around these points can be parametrized in terms of the Fermi velocity, $v_{\rm F}$, and
the radius of curvature $R_0$ of the FS. 
The change in energy induced by a nonzero $b$ value (out-of-plane component)
has a positive contribution (energy cost), $\Delta E_1$, arising from the first two terms of Eq.~\eqref{eq:Ham_KLM_b_exp_new_frame}
and a negative contribution (energy gain), $\Delta E_2$, that arises from the last two terms of Eq.~\eqref{eq:Ham_KLM_b_exp_new_frame}.
By following the procedure described in~\cite{supp}, we obtain
\begin{align}
\Delta E_1 &= - \frac{\Delta k}{4 \pi^2} \frac{J^2b^2}{16 v_{\rm F}} \ln{ \left [ \frac{x + \sqrt{(\frac{J}{2 v_{\rm F}})^2 +x^2}}
{ k_0+ x + \sqrt{(\frac{J}{2 v_{\rm F}})^2 + (k_0+x)^2}} \right ]},
\nonumber \\
\Delta E_2 &=  \frac{\Delta k}{4 \pi^2} \frac{J^2b^2}{16 v_{\rm F}} \ln{ \left[ \frac{x + \sqrt{(\frac{Jb}{4 v_{\rm F}})^2  +x^2}}
{ k_0+ x + \sqrt{(\frac{Jb}{4 v_{\rm F}})^2 + (k_0+x)^2}} \right],}
\label{eper}
\end{align} 
where $\Delta k$ and $k_0$ define the circular rectangle of integration around the hot spots [see Fig.~\ref{fig:perturbation}(a)], 
and $x={\Delta k}^2/8 R_0$. 
The result in Eq.~\eqref{eper} indicates that $\Delta E_1 + \Delta E_2 <0$ for $b \ll 1$, 
where the perturbative approach is valid. 
This reveals the origin of the noncoplanar magnetic ordering:
the single-$Q$ state with ${\bm q}={\bm Q}_1$ still has a high susceptibility for additional ${\bm Q}_2$ modulation.
It is also interesting to note that the energy gain $\Delta E_2$ becomes proportional to $b^2 \ln{b}$ for perfect nesting ($R_0 \to \infty$).

Figure~\ref{fig:perturbation}(c) shows the $\Omega(b)$ curve which is obtained from our perturbative approach 
after integrating over the ${\rm 1BZ}$ (not just around the hot spots)~\cite{supp}.
The energy gain shows a good agreement with the exact evaluation for the double-$Q$ state in Eq.~\eqref{eq:Spin_meron}, 
which can be computed by direct diagonalization of ${\mathcal H}$ 
for a commensurate wave number that makes the Hamiltonian block diagonalized.
The deviation at $b=0.1$ is $\mathcal{O}(10^{-8})= \mathcal{O}$($J^2b^6$, $J^4b^4$, $J^6b^2$), consistent with the expected accuracy of our perturbative expansion. 

\begin{figure}[t!]
	\includegraphics[width=8.0cm]{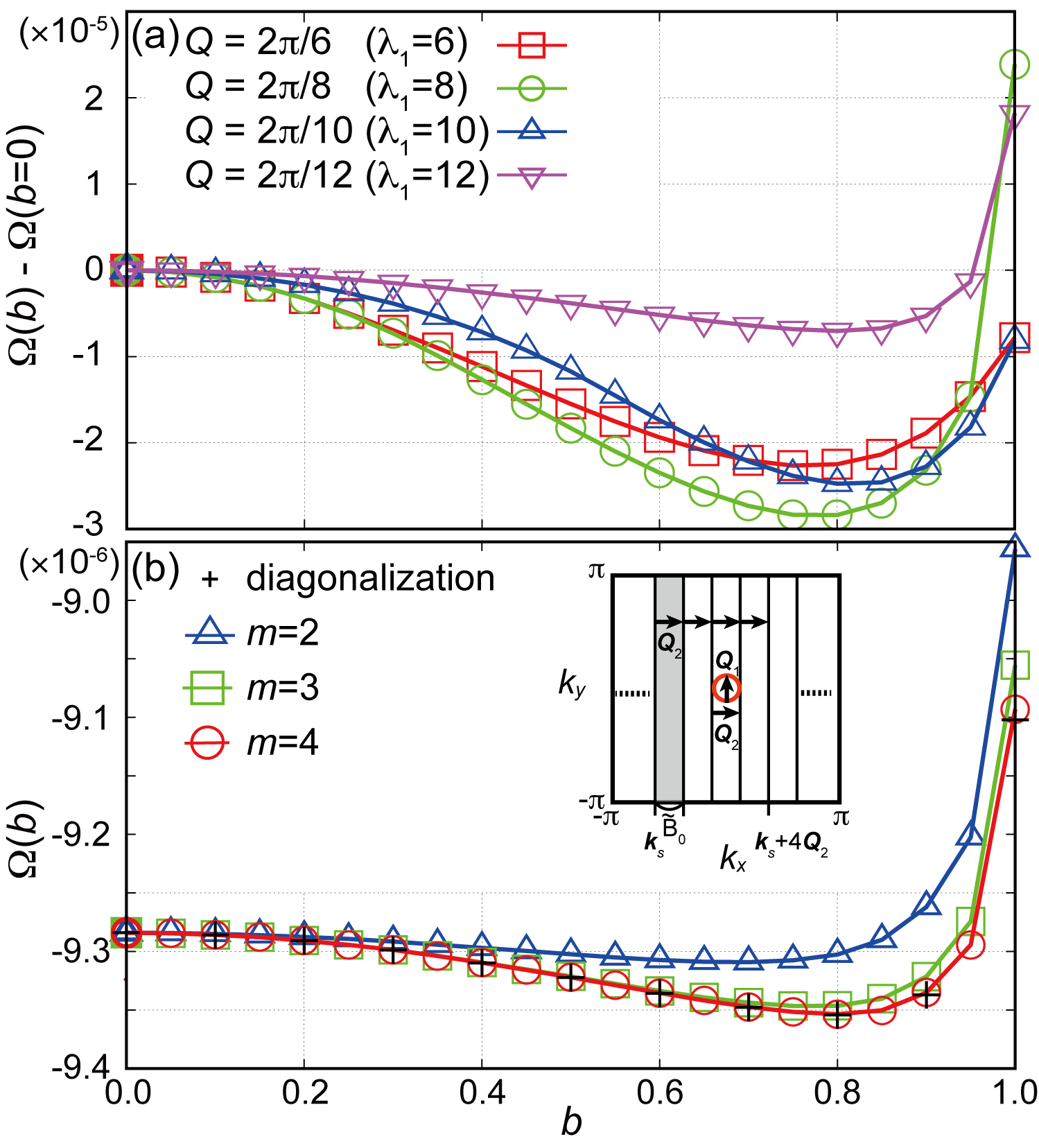}
	\caption{ \label{fig:variational}  (color online).
	 (a) Grand potential measured from that at $b=0$ as a function of $b$ 
	 in the $t_1$-model computed in a lattice of
	$N=960^2$ sites for several $\lambda_1$ and $\lambda_2=1$.
	(b) Convergence of the truncation approach. 
	 The triangles,  squares, and circles show the grand potential computed by including harmonics up to 
	 2${\bm Q}_2$,  3${\bm Q}_2$, and  4${\bm Q}_2$, 
	 respectively, for the $t_1$-model with $J=0.1$, $\mu=\mu^*(\lambda_1/\lambda_2=12)$, and $N=960^2$ sites.
	 Crosses represent the exact results obtained by direct diagonalization.
	 The shaded region in the inset shows the region in 1BZ where we take the sum of momentum 
	 for the truncation of scattering processes in the $m=4$ case~\cite{supp}. 
	 The red circle in the inset is a schematic of the FS. 
	 }
\end{figure}

Although the perturbative analysis thus far indicates the stability of the noncoplanar order in Eq.~\eqref{eq:Spin_meron} with $b\neq 0$, 
it does not allow us to determine the value $b = b_{\rm opt}$ that minimizes the grand potential $\Omega$. 
In the commensurate cases, i.e., the components of ${\bm Q}_{\nu}$ are $2\pi\lambda_2/\lambda_1$ 
($\lambda_1$ and $\lambda_2$ are integers), however, 
we can estimate $b_{\rm opt}$ by direct diagonalization of ${\mathcal H}$ as 
in Fig.~\ref{fig:perturbation}(c), at least, for not so large $\lambda_1$. 
The representative examples are shown in Fig.~\ref{fig:variational}(a) for $t_3=0$ ($t_1$-model):  ordering vectors are $\pm\bm{Q}_1=\pm 2 \pi (\lambda_2/\lambda_1, 0)$ and $\pm\bm{Q}_2=\pm 2 \pi (0, \lambda_2/\lambda_1)$ for $\mu=\mu^*(\lambda_1/\lambda_2)=-2-2\cos(\pi \lambda_2/\lambda_1)$. 
$b_{\rm opt}$ is close to one for all the values of $\lambda_2$, as the original perturbative analysis 
based on Eq.~\eqref{eq:del_energy_Jb_direct_simple} implies. 

To extend the computation of $b_{\rm opt}$ for arbitrary ${\bm Q}_{\nu}$, 
we once again work in the basis of quasi-particle operators that diagonalize  
${\cal H}$ for the ${\bm Q}_1$-helical ordering used in Eq.~(\ref{eq:Ham_KLM_b_exp_new_frame})~\cite{supp}. 
In this  basis, the ${\bm Q}_2$ component induces scattering processes between different harmonics: 
${\bm k} \to {\bm k}+ m {\bm Q}_2$, with $m$ being an integer.
These processes arise from the Fourier transform of $\sqrt{(1- b^2)+b^2\cos^2 \bm{Q}_2\cdot \bm{r}_i}$ in Eq.~(\ref{eq:Spin_meron}).
As shown in~\cite{supp}, the grand potential can be well approximated by truncating the harmonics at a certain $m$.  
This is demonstrated for a commensurate case in the $t_1$-model in Fig.~\ref{fig:variational}(b).
Comparison with direct diagonalization reveals that $b_{\rm opt}$ can be well estimated 
by keeping wave vectors up to the fourth harmonic, $4{\bm Q}_2$.
This truncation technique enables us to compute the grand potential in the generic case, even for incommensurate ordering vectors.

\begin{figure}[t]
	 \includegraphics[width=8.0cm]{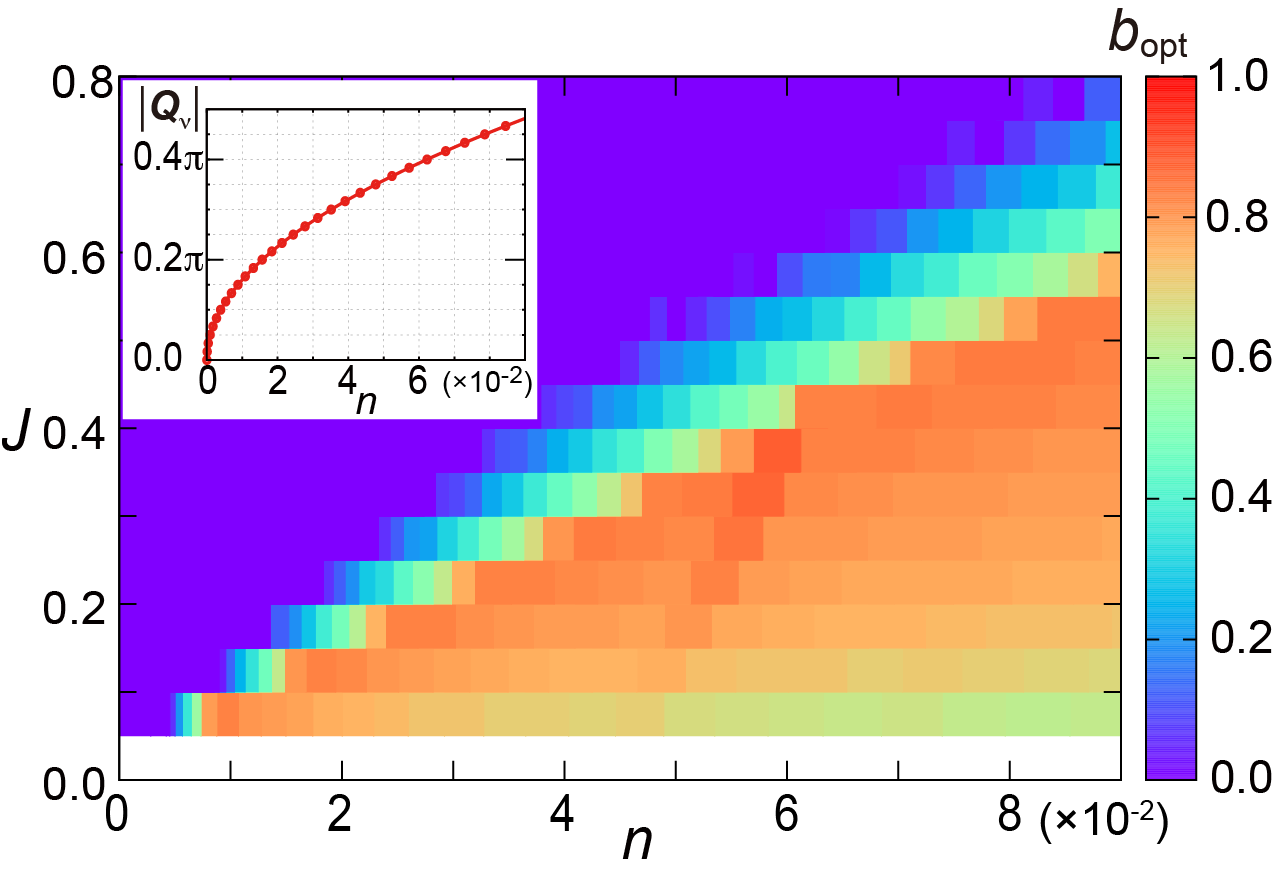}
	 \caption{ \label{fig:optbb} (color online).
	  Optimal degree of noncoplanarity $b$ in the $t_1$-model as a function of electron filling fraction $n$ and Hund coupling $J$, obtained by truncation of scattering processes at $m=4$.
	 Data below $J=0.05$ is not plotted due to numerical inaccuracy.
	 The inset shows corresponding wave vectors $|\bm{Q}_\nu|$ that we select to maximize susceptibility $\chi^0_{\bm q}$.}
\end{figure}

We now apply the above truncation technique to calculate the phase diagram of the $t_1$-model.
We use a sequence of ordering vectors $\bm{Q}_1=2 \pi (\lambda_2/\lambda_1, 0)$ 
and $\bm{Q}_2=2 \pi (0, \lambda_2/\lambda_1)$ where $\lambda_1=120$ and
$\lambda_2=1$, $2$, $\cdots$, $120$. For each $\bm{Q}_\nu$, 
we identify the chemical potential $\mu$ that minimizes the RKKY-level energy.
Having thus fixed $\mu$ and $\bm{Q}_\nu$, we estimate $b_{\rm opt}$ as a function of $J$. 
Figure~\ref{fig:optbb} shows our result in the ensemble of fixed electron filling 
$n =(1/N)\sum_{i\sigma}\langle c_{i\sigma}^\dagger c_{i\sigma}^{\;}\rangle$.
Note that $|b_{\rm opt}| \sim 1$ in the weak-coupling regime, consistent with our previous analysis. 
The inset of Fig.~\ref{fig:optbb} demonstrates that the length scale $1/|{\bm Q}_\nu|$ of vortices diverges as $n \rightarrow 0$.

In summary, we have demonstrated that frustration, here induced by multiple global maxima in the bare magnetic susceptibility, 
naturally leads to noncoplanar magnetic orderings in itinerant magnets. 
Whenever the ordering wave vectors connect {\it independent} pieces of the FS, 
the coplanar single-$Q$ magnetic ordering becomes unstable towards the generation of other ${\bm q}$ components
that gap out the corresponding (symmetrically related) pieces of the FS. 
In particular, for large magnetic moment systems (near the classical $S \to \infty$ limit), the additional
components appear through an out of plane modulation of the helical state in order to minimize the amplitude of the higher-harmonic components required by the constraint $|{\bm S}_i|=1$. 

For the square lattice considered here, this mechanism leads
to the stabilization of vortex crystals with spin scalar chiral stripes. 
It is important to note that noncoplanar magnetic orderings have been recently observed in 2D layers 
of $3d$ metals deposited on a nonmagnetic metallic surface~\cite{Heinze11,Yoshida12}.
Moreover, first-principles calculations for these systems~\cite{Heinze11,Yoshida12,Kurz01} indicate that these noncoplanar orderings arise from rather strong effective four-spin interactions. 
Our results unveil the generic origin of these effective interactions and explain why they are so ubiquitous in frustrated itinerant magnets. 
More exotic structures may appear in higher symmetry 3D lattices in which the 
multiple-ordering wave vectors do not lie on the same plane~\cite{Wang2015,Tanigaki2015}.
Indeed, the 3D frustrated itinerant magnet SrFeO$_3$~\cite{Ishiwata11} is a strong candidate to exhibit multiple-$Q$ magnetic ordering based on the mechanism described in this work. 

\begin{acknowledgments}
We thank  S. Bl\"ugel, 
Y. Kato, S. Kumar, S.-Z. Lin, T. Misawa, M. Udagawa, A. Uehara, Y. Yamaji, and T. Ziman  for fruitful discussions.
R.O. is supported by the Japan Society for the Promotion of Science through a research fellowship for young scientists and the Program for Leading Graduate Schools (ALPS). 
R.O. also acknowledges the CNLS summer student program at LANL.
This reaserch is supported  by Grants-in-Aid for Scientific Research (No. 24340076), the Strategic Programs for Innovative Research (SPIRE), MEXT, and the Computational Materials Science Initiative (CMSI), Japan. Work at LANL was carried out under the auspices of the U.S. DOE contract No. DE-AC52-06NA25396 through the LDRD program.
\end{acknowledgments}

\nocite{*}

\bibliography{reference}

\begin{thebibliography}{32}
\bibitem[{sma()}]{smallpoint}
\bibinfo{note}{By low-symmetry wave vectors, we mean those whose small point
  group of the wave vectors has no more than two elements.}
\bibitem[{num()}]{numeric}
\bibinfo{note}{Details of the numerical simulations will be reported
  elsewhere.}
\bibitem[{loc()}]{localchi}
\bibinfo{note}{$\chi_i \equiv (\chi_{ijk} + \chi_{ijl} + \chi_{ilk} +
  \chi_{jlk}$)/4, where ${\bm r}_j={\bm r}_i + {\bm e}_x$, ${\bm r}_k={\bm r}_i
  + {\bm e}_y$, and ${\bm r}_l={\bm r}_i + {\bm e}_x + {\bm e}_y$ [${\bm
  e}_{x(y)}$ is the unit vector parallel to ${\hat {\bm x}}({\hat {\bm y}})$].}
\bibitem[{2or()}]{2order}
\bibinfo{note}{This second-order treatment includes all the Feynman diagrams up
  to fourth order in the original basis [see Eq.~(4)] that contribute to the
  total energy of the spin configuration given by Eq.~(3), plus all the other
  diagrams that are generated by multiple-${\bm Q}_1$ scattering processes.}
\end{thebibliography}

\end{document}